\documentclass[slac_one]{revtex4}
\usepackage{graphicx}
\usepackage{fancyhdr}
\usepackage{xspace}

\newcommand{\piz} {\ensuremath{\pi^0}\xspace}
\newcommand{\pT} {\ensuremath{p_\mathrm{T}}\xspace}
\newcommand{\sNN} {\ensuremath{\sqrt{s_{NN}}}\xspace}
\newcommand{\dphi} {\ensuremath{\Delta\phi}\xspace}
\newcommand{\deta} {\ensuremath{\Delta\eta}\xspace}
\newcommand{\RAA} {\ensuremath{R_{AA}}\xspace}
\newcommand{\qhat} {\ensuremath{\left<\hat{q}\right>}\xspace}

\begin{document}

%Title of paper
\title{Measurement of Large Transverse Momentum Hadrons and
Constraints on Medium Opacity Parameters} %% Paper title goes here

% Repeat the \author .. \affiliation  etc. as needed
%
% \affiliation command applies to all authors since the last
% \affiliation command. The \affiliation command should follow the
% other information

\author{D. Winter (for the PHENIX collaboration)}
\affiliation{Columbia University, New York, NY, USA}

\begin{abstract}
The PHENIX experiment has measured {\piz}s in Au+Au collisions at
$\sqrt{s_{NN}}$ = 200 GeV, with good statistics for transverse momentum, \pT,
up to 20 GeV/c. A fivefold suppression is found, which is essentially constant
for $5<\pT<20$ GeV/c.  While the production of high \pT pions in high energy
p+p collisions is well described in the framework of perturbative QCD,
production in ultra-relativistic heavy ion collisions involves additional
degrees of freedom related the opacity of the deconfined medium produced in
the collisions. The uncertainties of the latest measurement are small enough
to constrain model-dependent opacity-related parameters, such as the transport
coefficient of the medium, $\hat{q}$ or initial gluon density $dN^g/dy$ at the
level of $\pm$ 20-25\% (one standard deviation), by applying a statistical
method proposed. From the dependence of the suppression level to the number of
participant nucleons, we have also confirmed that the energy loss increases as
\pT of partons increases.  The latest PHENIX result on high transverse
momentum hadrons are presented, and the property of the medium created in the
collisions is discussed in detail.
\end{abstract}

%\maketitle must follow title, authors, abstract
\maketitle

\thispagestyle{fancy}

% body of paper here - Use proper section commands
% References should be done using the \cite, \ref, and \label commands
% Put \label in argument of \section for cross-referencing
%\section{\label{}}

\section{INTRODUCTION} % Section title should be in all capitals.

It has now been established that novel form of matter is created in
the ultra-relativistic heavy-ion collisions at the Relativistic Heavy
Ion Collider at Brookhaven National Laboratory~\cite{Adcox:2004mh}.
Hard scattering of individual partons is a particularly important
probe of the evolution of the medium created in these collisions, as
it occurs at early stages of the collision and produces high-\pT
observables.  The observation of jet
quenching~\cite{Adler:2005ee,Adare:2008cq} has motivated a great deal
of theoretical and experimental work in an attempt to understand the
source of this effect.  Suppression exhibited in measured spectra is
currently attributed to energy loss of fast moving partons as they
traverse the medium, prior to fragmentation into observed particles.
There are two sources of energy loss considered responsible: coherent
medium-dependent gluon bremsstrahlung and collisional energy loss.

A number of models have been used to predict the energy loss seen in
heavy-ion collisions, representing a variety of implementatons of the
collision geometry and the degree to which they represent the two
sources of energy loss (for example, see~\cite{Adare:2008cg} and
references therein).  In most cases, these models can be characterized
by a single input parameter.  The input parameter typically describes
the opacity of the medium; the more opaque the medium, the greater the
energy loss and therefore the greater the suppression observed.  We
report on recent PHENIX measurements of high-\pT hadrons, including
the impact the results have on constraining several popular models.

%% In addition to the data presented, we report on analysis using two models,
%% GLV and PQM, used to compare with the data.  The reader is invited to
%% explore the references, specifically~\cite{Adare:2008cg} and references therein,
%% for a description of several others.

\section{PHENIX MEASUREMENT OF HIGH PT HADRONS}

The PHENIX detector~\cite{Morrison:1998qu} is used to measure {\piz}s via
their decay to photons using an electromagnetic calorimeter (EMC) that covers
$\left|\eta\right|<0.35$. The EMC consists of eight sectors, each covering
$\pi/8$ in azimuth, using two detector technologies: Pb-scintillating (PbSc)
sampling calorimeter (six sectors) and Pb-glass (PbGl) Cherenkov calorimeter
(two sectors).  The cell size $\dphi \times \deta$ of the calorimeter is
$0.011^2$ and $0.0075^2$ for the PbSc and PbGl, respectively.  For heavy-ion
collisions, the centrality (degree of overlap between nuclei) of the collision
is measured using the correlation between the charged multiplicity measured in
the Beam-Beam Counters (BBCs) and the energy deposited in the zero-degree
calorimeters (ZDCs).

\begin{figure}
\begin{center}
\includegraphics[height=0.4\textheight]{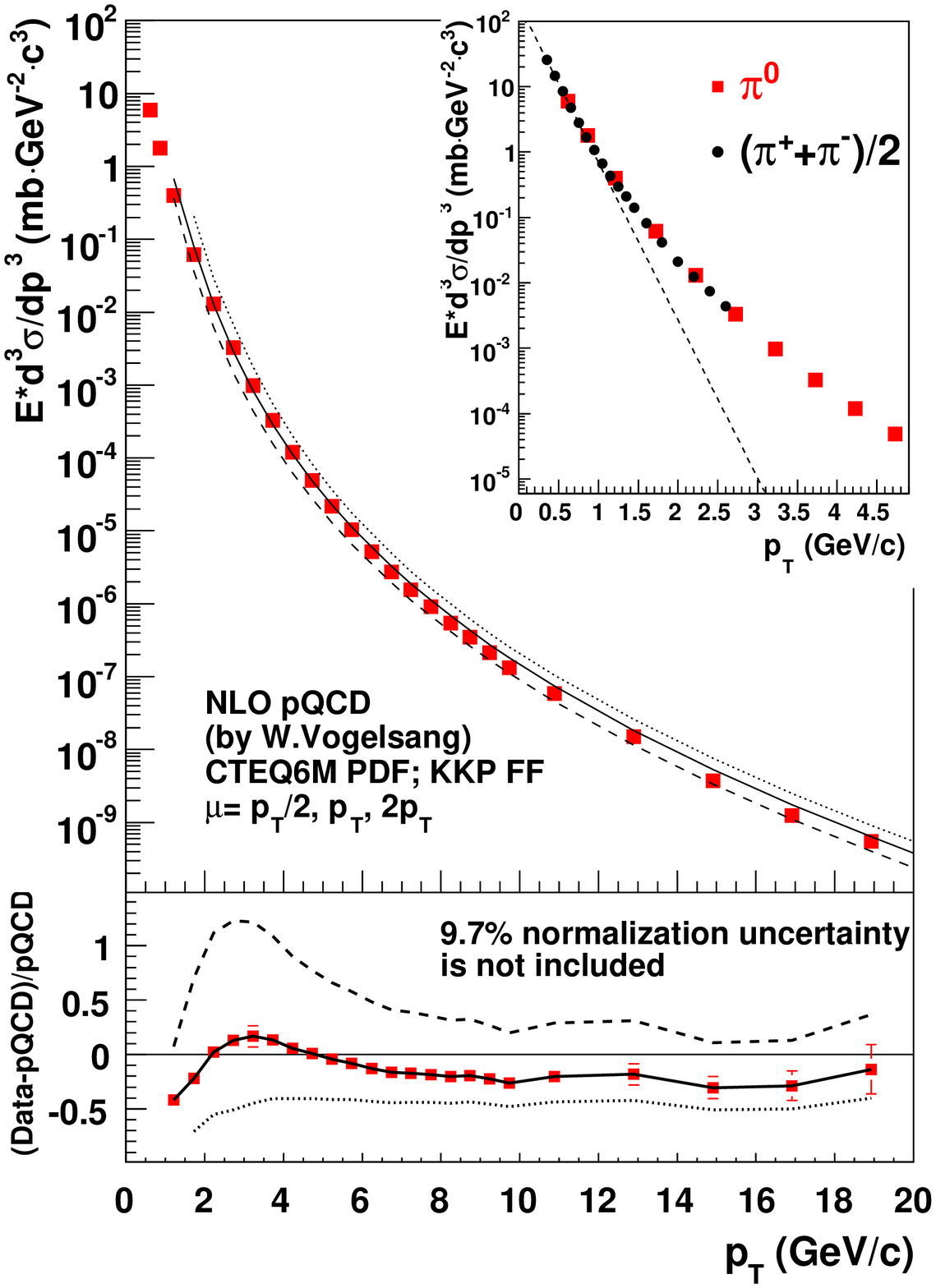}
\includegraphics[height=0.35\textheight]{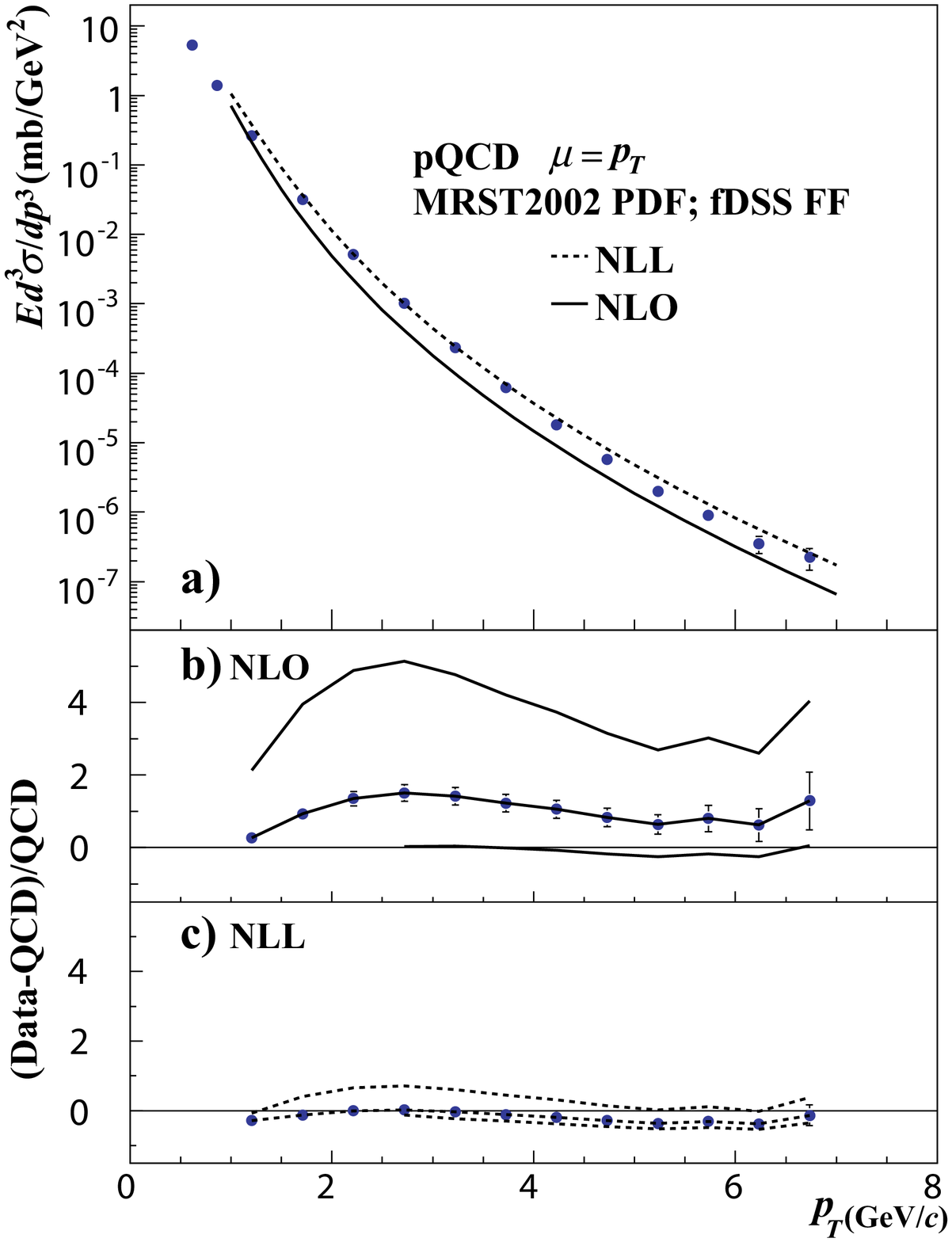}
\end{center}
\caption{Measured \piz spectra and QCD preditions. LEFT: Neutral pion
  cross-section measured in p+P collisions at $\sNN=200$ GeV/c.  For
  comparison, the cross-section for charged pions is included.  RIGHT:
  Cross-section as measured in p+p collisions at $\sNN=62.4$ GeV/c}
\label{fig:pp_xs}
\end{figure}

We have measured the \piz spectrum in p+p collisions at \sNN = 200
GeV~\cite{Adare:2007dg} and 62.4 GeV~\cite{Adare:2008qb}, as shown in
Figure~\ref{fig:pp_xs}.  The measured cross-sections show quite good agreement
with pQCD calculations, which provides an important baseline for comparison
with heavy ion collisions as well as establishes the fact that we indeed
observe hard-scattering in RHIC collisions.

\begin{figure}
\begin{center}
\includegraphics[width=0.48\textwidth]{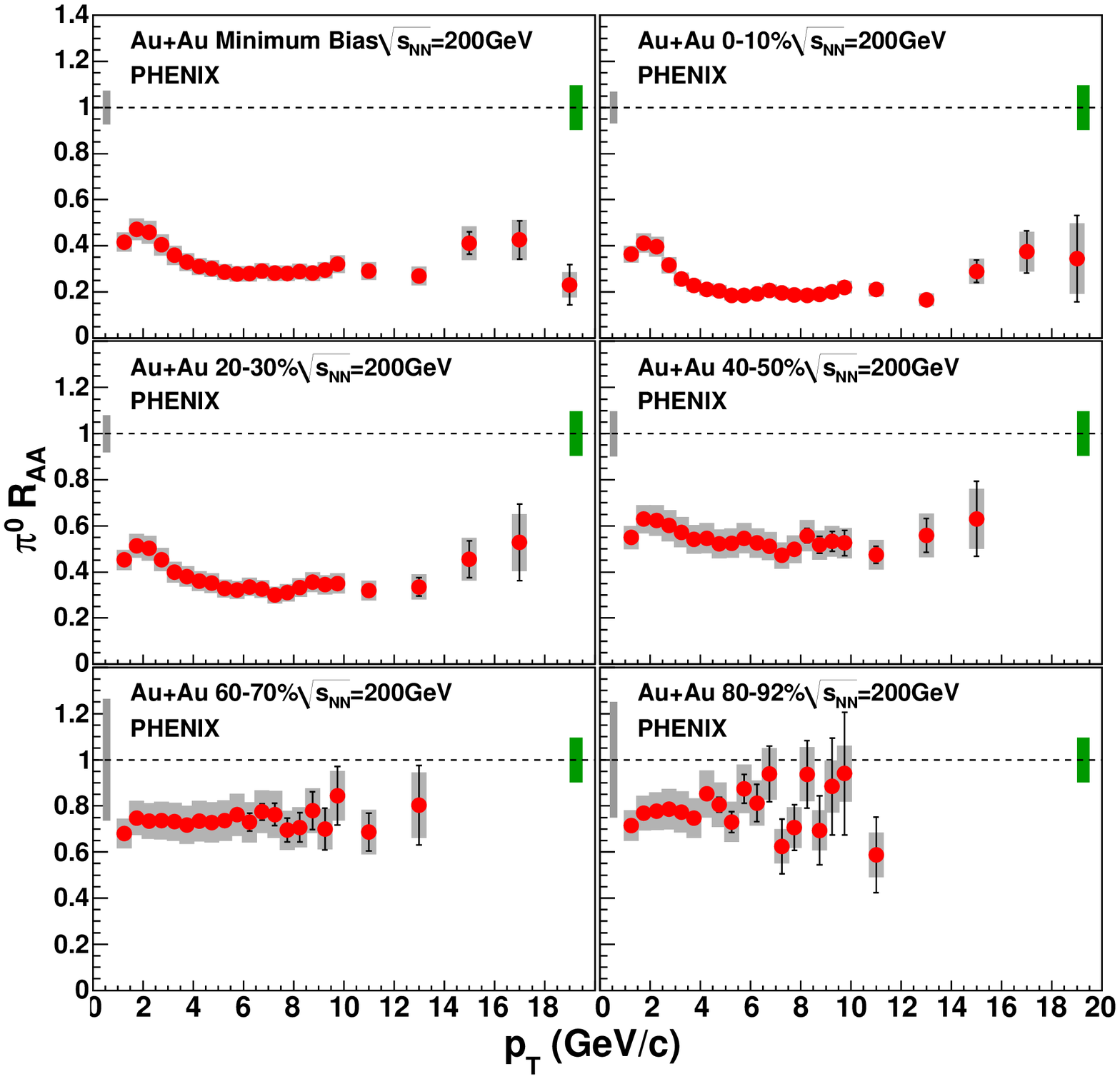}
\includegraphics[width=0.48\textwidth]{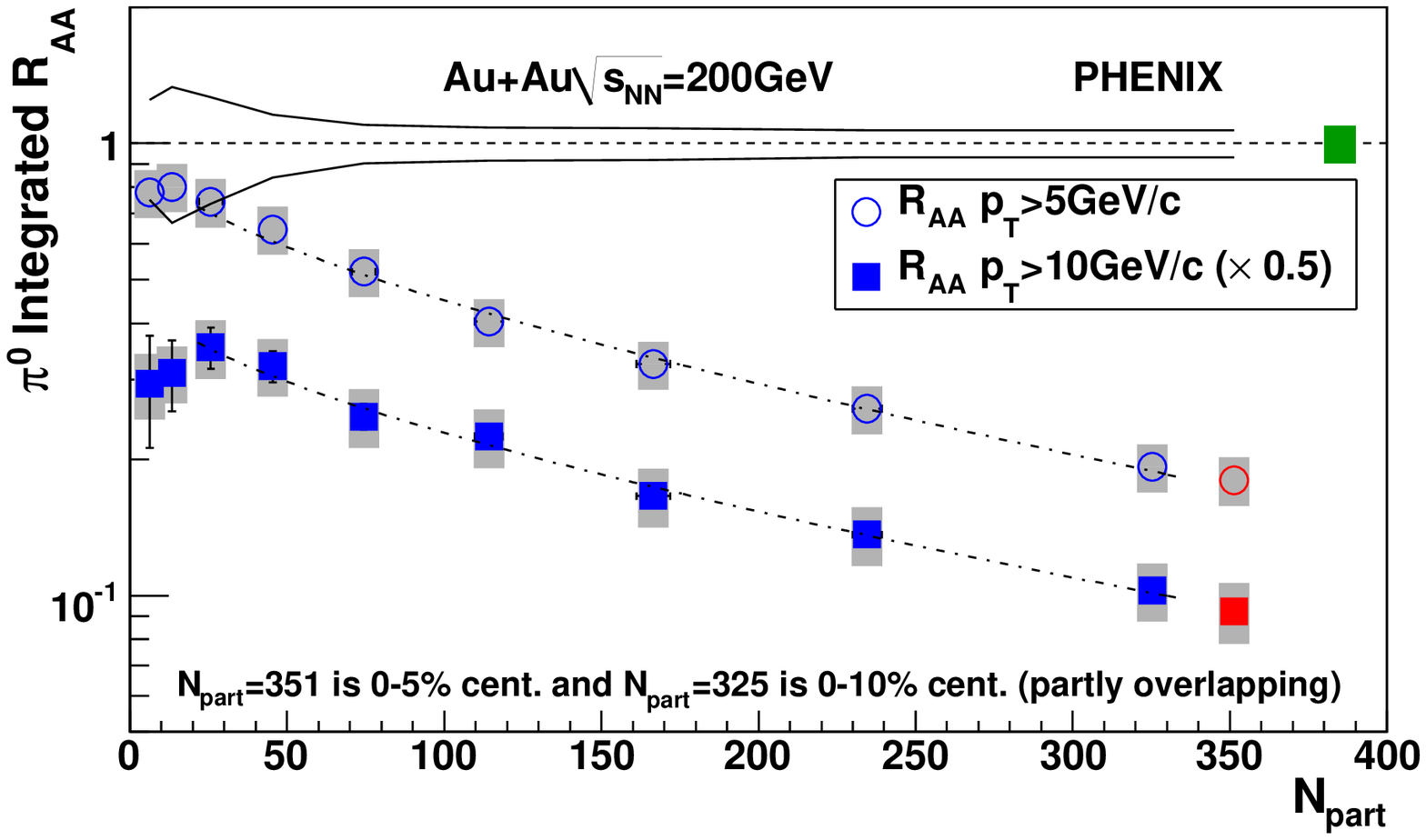}
\end{center}
\caption{LEFT: \piz \RAA versus \pT as a function of centrality, for
  Au+Au collisions at $\sNN=200$ GeV/c.  Statistical errors are
  indicated with error bars, point-to-point systematic errors as boxes
  around the pointes, and global systematics as error boxes at
  $\RAA=1$.  The box on the left is the error due to calculation of
  the number of binary collisions, and the box on the right is the
  error due to the normalization of the p+p reference spectrum.
  RIGHT: Integrated \piz \RAA as a function of the number of
  participants.  The fits are described in the text.}
\label{fig:pi0_RAA}
\end{figure}

Neutral pion spectra have been measured in both Au+Au and Cu+Cu
collisions, as a function of centrality of the
collision\cite{Adare:2008qa,Adare:2008cx}.  When this spectra are
compared to that of p+p collisions scaled by the appropriate number of
binary parton-parton collisions, an important suppression pattern is
revealed.  For peripheral collisions, the measured spectrum is similar
to the p+p spectrum, implying that the heavy-ion collision is a
superposition of individual hadron collisions.  However, in the more
central collisions we observe suppression of produced particles as
large as a factor of five, and approximately independent of \pT for
the high-\pT regime.  The suppression seen in nuclear collisions is
characterized by the nuclear modification factor, \RAA:
\begin{equation}
\RAA = \frac{1/N_{evts}d^2N/dyd\pT}{\left<T_{AB}\right>d^2\sigma_{pp}/dyd\pT}
\end{equation}
where $\sigma_{pp}$ is the production cross-section in p+p collisions,
and $\left<T_{AB}\right>$ is the nuclear thickness function averaged
over a range of impact parameters in the particular centrality,
calculated with a Glauber Model The \RAA for neutral pions in Au+Au
collisions has been measured by PHENIX most recently in
2004~\cite{Adare:2008qa} and is shown in Figure~\ref{fig:pi0_RAA}.

In order to use the \RAA to shed light on energy loss in the medium,
we note that for $\pT>5.0$~GeV/c both the p+p and Au+Au spectra behave
as a power law.  This allows a reinterpretation of the suppression as
an effective fractional energy loss, $S_{loss} = 1-\RAA^{1/(n-2)}$.
$S_{loss}$ is expected to be proportional to a power of the number of
particpant nucleons, $N_{part}^a$.  We fit the $N_{part}$ dependence
of the \RAA integrated over two different \pT ranges with the function
$\RAA=(1-S_0N_{part}^a)^{n-2}$.  The result is shown in the right
panel of Figure~\ref{fig:pi0_RAA}.  The \RAA versus centrality does
not appear to saturate, and the fit gives us $a = 0.58\pm 0.07
(0.56\pm0.07)$, $S_0=(8.3\pm3.3)\times10^{-3}
((9.2\pm4.9)\times10^{-3})$ for $\pT>5$~GeV/c ($\pT>10$~GeV/c).  Both
the Gyulassy-Levai-Vitev (GLV) model and parton quenching model (PQM)
predict $a\simeq2/3$.  All this evidence points to the implication
that the energy loss experienced by partons moving through the medium
increases with \pT.

\section{QUANTITATIVE ANALYSIS}

A new statistical analysis has been proposed to help constrain models,
using a complete description of all experimental
uncertainties~\cite{Adare:2008cg}.  Given a model prediction that
depends on a single input parameter $p$, $\mu(p)$, a fit is performed
on the data by varying $p$ and the most likely range of values are
determined at the one and two $\sigma$ levels.  This fit involves
careful categorization of uncertainities: Type A, point-to-point
uncorrelated (statistical) errors; Type B, point-to-point correlated
errors; and Type C, globally correlated errors.  We then seek to
minimize the $\tilde{\chi}^2$ statistic:
\begin{eqnarray}
\tilde{\chi}^2(\epsilon_a,\epsilon_c,p) & = & \left[\sum_{i=1}^{n}
  \frac{(y_i^2+\epsilon_b\sigma_{b_i}+\epsilon_c y_i\sigma_c -
  \mu_i(p))^2}{\tilde{\sigma_i}^2}+\epsilon^2_b+\epsilon_c^2\right] \\
  \tilde{\sigma_i}^2 & = & \sigma_i
  (y_i^2+\epsilon_b\sigma_{b_i}+\epsilon_c y_i\sigma_c)/y_i
\end{eqnarray}
where $\epsilon_b$ and $\epsilon_c$ are the fractions of the Type B
and C errors that displace all points together.  It is important to
note that this procedure accounts for only the experimental
uncertainties. 

\begin{figure}
\begin{center}
\includegraphics[width=0.9\textwidth]{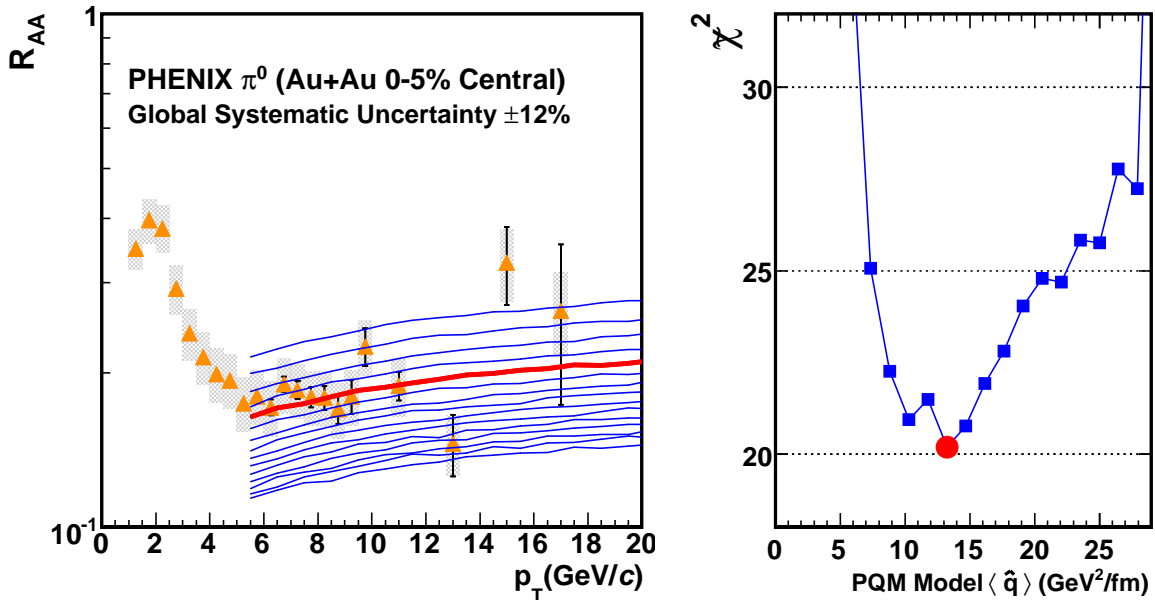}
\end{center}
\caption{Fits of PQM to \piz \RAA for 0-5\% central events in Au+Au
  collisions at $\sNN=200$ GeV/c.  LEFT: Neutral pion \RAA, shown with
  the resulting PQM predictions as \qhat is varied.  RIGHT:
  Chi-squared of the PQM fit as a function of \qhat.  The red dot
  indicates the minimum}
\label{fig:centRAA_pqmfit}
\end{figure}

As an example, this analysis has been applied to the 0-5\% most
central \RAA measurements using PQM~\cite{Adare:2008cg}.  PQM is a
Monte Carlo model based on the quenching weights from the BDMPS model.
It is characterized by a single parameter, \qhat, which quantifies the
average transverse momentum transferred from the medium to the parton
per unit mean free path.  The result of varying \qhat and fitting to
the \RAA data is shown in Figure~\ref{fig:centRAA_pqmfit}.  The data
constrains \qhat as $13.2$ $^{+2.1}_{-3.2}$ $^{+6.3}_{-5.2}$
GeV$^2$/fm at the one and two $\sigma$ levels.  Similar analysis has
been performed on several other popular energy-loss models, producing
similar constraints on their input parameters~\cite{Adare:2008cg}.

\section{SUMMARY}

PHENIX has measured high-\pT ($1<\pT<20$~GeV/c) hadrons in Au+Au
collisions at $\sNN$ = 200 GeV.  In the most central collisions and
above $\pT\simeq 5$~GeV/c, we observe a suppression of hadrons of a
factor of approximately five.  This suppression remains constant out
to the highest accessible \pT.  The pattern of suppression as seen in
the \RAA measurement suggests that the energy loss suffered by the
partons produced in these collisions is proportional to \pT: we find
that the integrated \RAA does not saturate with the number of
participants, the data are consistent with an $S_{loss} \propto
N_{part}^a$, and the high-\pT spectra follows the same power law from
p+p to Au+Au (with $n\simeq 8.1$).  Most importantly, the precision of
our \RAA measurement allows for the application of a quantitative
approach to finding constraints to existing models.  Incorporating all
experimental uncertainties into this analysis, we are able to
constrain the single input parameters for several popular models to
10-20\% at the one standard deviation level.

\bibliographystyle{unsrt}
\bibliography{ICHEP2008_Winter_Proceedings}

\begin{thebibliography}{1}

\bibitem{Adcox:2004mh}
K.~Adcox et~al.
\newblock {Formation of dense partonic matter in relativistic nucleus nucleus
  collisions at RHIC: Experimental evaluation by the PHENIX collaboration}.
\newblock {\em Nucl. Phys.}, A757:184--283, 2005.

\bibitem{Adler:2005ee}
Stephen~Scott Adler et~al.
\newblock {Modifications to di-jet hadron pair correlations in Au + Au
  collisions at $\sqrt{s_{NN}}$ = 200-GeV}.
\newblock {\em Phys. Rev. Lett.}, 97:052301, 2006.

\bibitem{Adare:2008cq}
A.~Adare et~al.
\newblock {Dihadron azimuthal correlations in Au+Au collisions at
  $\sqrt{s_{NN}}$ = 200 GeV}.
\newblock {\em Phys. Rev.}, C78:014901, 2008.

\bibitem{Adare:2008cg}
A.~Adare et~al.
\newblock {Quantitative Constraints on the Opacity of Hot Partonic Matter from
  Semi-Inclusive Single High Transverse Momentum Pion Suppression in Au+Au
  collisions at $\sqrt{s_{NN}}$ = 200 GeV}.
\newblock {\em Phys. Rev.}, C77:064907, 2008.

\bibitem{Morrison:1998qu}
D.~P. Morrison et~al.
\newblock {The PHENIX experiment at RHIC}.
\newblock {\em Nucl. Phys.}, A638:565--570, 1998.

\bibitem{Adare:2007dg}
A.~Adare et~al.
\newblock {Inclusive cross section and double helicity asymmetry for $\pi^0$
  production in p+p collisions at $\sqrt{s}$ = 200 GeV: Implications for the
  polarized gluon distribution in the proton}.
\newblock {\em Phys. Rev.}, D76:051106, 2007.

\bibitem{Adare:2008qb}
A.~Adare et~al.
\newblock {Inclusive cross section and double helicity asymmetry for $pi^0$
  production in p+p collisions at $\sqrt{s}$ = 62.4 GeV}.
\newblock 2008.
\newblock arXiv:hep-ex/0810.0701.

\bibitem{Adare:2008qa}
A.~Adare et~al.
\newblock {Suppression pattern of neutral pions at high transverse momentum in
  Au+Au collisions at $\sqrt{s_{NN}}$ = 200 GeV and constraints on medium
  transport coefficients}.
\newblock 2008.
\newblock Accepted for publication in Phys. Rev. Lett.

\bibitem{Adare:2008cx}
A.~Adare et~al.
\newblock {Energy dependence of pi-zero production in Cu+Cu collisions at
  $\sqrt{s_{NN}}$ = 22.4, 62.4, and 200 GeV}.
\newblock 2008.

\end{thebibliography}

\end{document}